\documentclass[reprint,groupedaddress,superscriptaddress,amsmath,amssymb,prl]{revtex4-2}

\usepackage[utf8]{inputenc}
\usepackage{graphicx}
\usepackage{dcolumn}
\usepackage{bm}
\usepackage[%
    colorlinks=true,
    urlcolor=magenta,
    linkcolor=red,
    citecolor=cyan
]{hyperref}
\usepackage[mathlines]{lineno}
\usepackage{siunitx}
\usepackage{amsmath}
\usepackage{physics}
\usepackage{multirow}
\usepackage[nodisplayskipstretch]{setspace}
\newcommand*{\xdash}[1][1em]{\rule[0.5ex]{#1}{0.55pt}}
\usepackage[most]{tcolorbox}
\usepackage{xcolor}
\usepackage{soul}

\begin{document}

\title{Singular Topological Edge States in Locally Resonant Metamaterials}

 \author{Yeongtae Jang}\email{These authors contributed equally to this work.}
     \affiliation{%
     Department of Mechanical Engineering, Pohang University of Science and Technology (POSTECH), Pohang 37673, Republic of Korea}
     \email{jrsho@postech.ac.kr}
 \author{Seokwoo Kim}\email{These authors contributed equally to this work.}
      \affiliation{%
      Department of Mechanical Engineering, Pohang University of Science and Technology (POSTECH), Pohang 37673, Republic of Korea}
  \author{Eunho Kim}\email{eunhokim@jbnu.ac.kr}
      \affiliation{%
        Division of Mechanical System Engineering, Jeonbuk National University, Jeonju, 54896, Republic of Korea}
        \email{eunhokim@jbnu.ac.kr}
      \affiliation{%
     Department of JBNU-KIST Industry-Academia Convergence Research, Jeonju, 54896, Republic of Korea}
 \author{Junsuk Rho}\email{jrsho@postech.ac.kr}
     \affiliation{%
     Department of Mechanical Engineering, Pohang University of Science and Technology (POSTECH), Pohang 37673, Republic of Korea}
    \email{jrsho@postech.ac.kr}
     \affiliation{%
     Department of Chemical Engineering, Pohang University of Science and Technology (POSTECH), Pohang 37673, Republic of Korea}
    \affiliation{%
    Department of Electrical Engineering, Pohang University of Science and Technology (POSTECH), Pohang 37673, Republic of Korea}
     \affiliation{%
     POSCO-POSTECH-RIST Convergence Research Center for Flat Optics and Metaphotonics, Pohang 37673, Korea}

\date{\today}

\begin{abstract}
Band topology has emerged as a novel tool for material design across various domains, including photonic and phononic systems, and metamaterials. 
A prominent model for band topology is the Su-Schrieffer-Heeger (SSH) chain, which reveals topological in-gap states within Bragg-type gaps (BG) formed by periodic modification.
Apart from classical BGs, another mechanism for bandgap formation in metamaterials involves strong coupling between local resonances and propagating waves, resulting in a local resonance-induced bandgap (LRG).
Previous studies have shown the challenge of topological edge state emergence within the LRG.
Here, we reveal that topological edge states can emerge within an LRG by achieving both topological phase and bandgap transitions simultaneously.
We describe this using a model of inversion-symmetric extended SSH chains for locally resonant metamaterials.
Notably, this topological state can lead to highly localized modes, comparable to a subwavelength unit cell, when it emerges within the LRG.
We experimentally demonstrate distinct differences in topologically protected modes—highlighted by wave localization—between the BG and the LRG using locally resonant granule-based metamaterials.
Our findings suggest the scope of topological metamaterials may be extended via their bandgap nature.
\end{abstract}
\maketitle
\section{Introduction}
\begin{figure*}
\centering
    \includegraphics [width=0.98\textwidth]{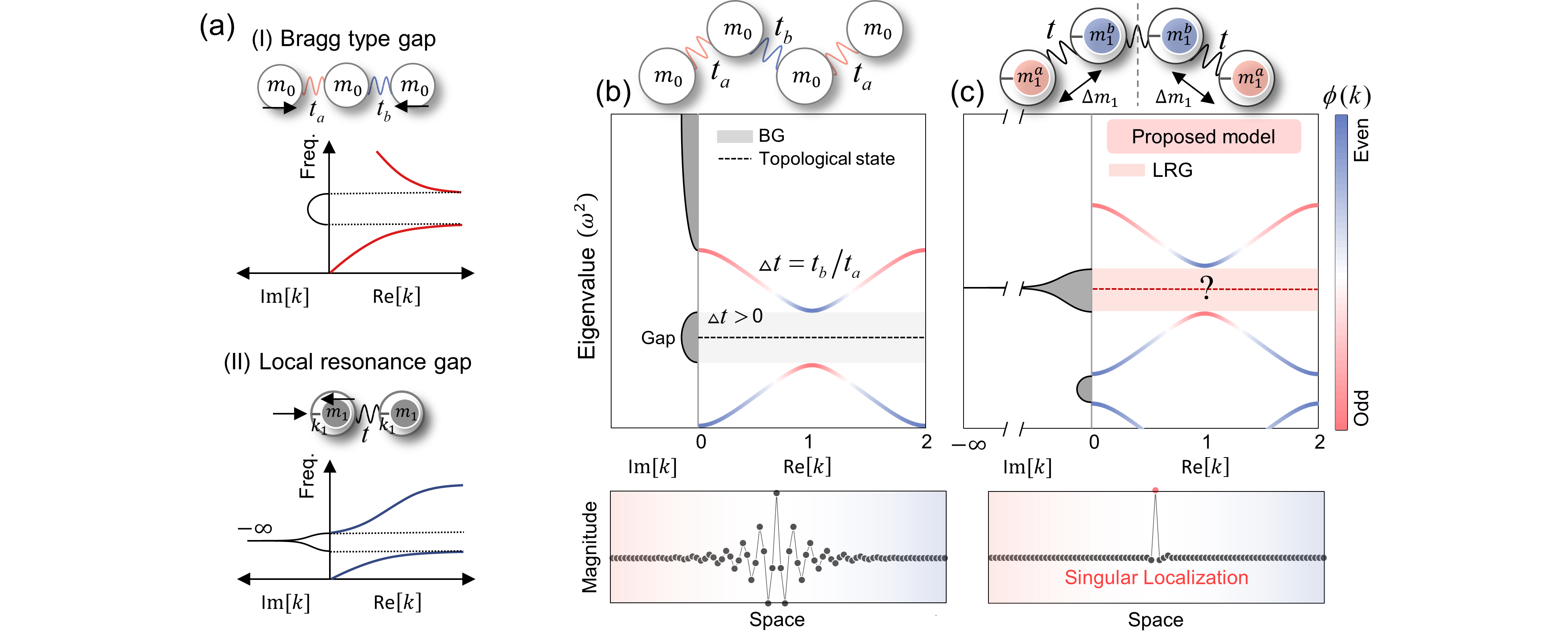}
    \caption{\textbf{Singular topological edge state.} (a) Classification of two types of bandgaps: Bragg-type gap and local resonance-type gap, based on their underlying physical principles. (b) Conventional Su-Schrieffer-Heeger (SSH) chain configured with dimerized hopping (stiffness coefficient). The right is the real part and the left is the imaginary part of the band structures respectively. The topological edge states emerge inside the Bragg-type gap (gray-shaded area). 
    At the bottom is the mode profile of their topological edge states. (c) Our proposed model system of inversion-symmetric extended SSH chains for locally resonant metamaterials, characterized by bi-local resonance frequencies. In this scheme, a topological edge state can emerge inside the strong local resonance gap (red-shaded area), leading to highly localized modes comparable to a subwavelength unit cell.}
    \label{fig1}
\end{figure*}

Band topology, a landmark of 20th-century physics~\cite{Hasan2010}, has emerged as a novel design strategy for photonic crystals, phononic crystals, and various metamaterials~\cite{Lu_2014,Ozawa_2019,Xue_2022}.
The core idea of band topology is to characterize the band structure of an infinite bulk system and predict the behavior of its local boundaries in finite systems~\cite{Qi2011,Bansil2016}.
This bulk-boundary correspondence~\cite{Prodan_2016} has emerged as a new degree of freedom for designing materials with intriguing energy localization features on their corners, edges, and surfaces~\cite{S_sstrunk_2015,Nash_2015,Serra_Garcia_2018,Cha_2018,Wang_2015,Vila_2017,Xiao_2015,Kane_2013,Pal_2017,Mousavi_2015,Rocklin_2016,Chaunsali_2018}.
What makes this special is that the topological nature of the bulk ensures robustness to the boundaries against perturbation~\cite{Asb_th_2016}.
This characteristic of topology opens up potential applications in optical, phononic, and mechanical computing~\cite{Ozawa_2019,Yasuda_2021,Huber_2016}, energy harvesting~\cite{Huber_2016}, and signal processing applications, among others~\cite{Zangeneh_Nejad_2019,Pirie_2022}.
\\
\indent
The topological “in-gap” eigenstates of finite systems are determined by the specific symmetries and bandgap properties (i.e., the imaginary part of the wavevectors) assigned to them. 
Typically, bandgaps can be categorized into two types according to their underlying physical principles: “Bragg-type” bandgaps (BG) and “local resonance-type” bandgaps (LRG) [Fig.~\ref{fig1}(a)].
One of the most fundamental systems in topological band theory is the one-dimensional Su-Schrieffer-Heeger (SSH) dimer model~\cite{Su1979}, which exhibits BGs.
This system undergoes transitions between different topological phases when the inter- and intra-particle potentials are manipulated, leading to the emergence of edge states at the boundaries between distinct topologies [Fig.~\ref{fig1}(b)].
\\
\indent
Apart from the BG, the LRG is a key feature of metamaterials, enabling subwavelength functionalities and characterized by strong wave attenuation~\cite{Liu_2000,Cummer_2016,Ma_2016}.
This behavior is explained by solutions involving the imaginary part of the band structures.
Indeed, the dispersion character of the local resonant system is similar to that of polaritons—a strong coupling between the propagating waves within the host medium and the resonance of the unit cell~\cite{Kaina_2015, Yves_2017_1}.
To realize such resonance, diverse design principles have been proposed and successfully applied in the fields of photonics~\cite{Smith_2004,Zheludev_2012}, and phononics~\cite{Liu_2000,Cummer_2016,Ma_2016}.
\\
\indent
Building on this background, we asked the following question:~\emph{Can we achieve nontrivial topological states within LRGs}?~[Fig.~\ref{fig1}(c)].
Since the localization properties of in-gap states rely on the bandgap properties, the topological edge state within the LRG is expected to exhibit stronger localization compared to conventional BGs.
Indeed, by combining locally resonant metamaterials with band topology, previous studies have successfully demonstrated the emergence of topological states, ranging from locally resonant 1D SSH models)~\cite{Zhao_2018,Li_2020,Coutant_2022,Zhao_2021} to 2D crystalline-based metamaterials~\cite{Yves_2017_1,Chaunsali_2018,Foehr_2018,Lee_2019,Fan_2023,Zhang_2019}.
However, in most of these studies, the role of local resonance has primarily been to lower or tune the frequency of edge states. Therefore, while topological in-gap states can achieve subwavelength characteristics, they still depend on BGs.
To date, implementing nontrivial topological states within the framework of LRGs has remained challenging.
The primary reason is that LRGs exhibit singularities in their gaps where the effective parameters diverge, thus preventing the necessary gap closure for band inversion~\cite{Zhao_2018}.
\\
\indent
In this study, we reveal the fundamental mechanism for achieving a topological state in LRGs within subwavelength regimes.
Interestingly, the topological in-gap states can be located near singularities in the imaginary part of the wavenumber within the LRGs [see Fig.~\ref{fig1}(a)].
This demonstrates extreme wave localization characteristics on a scale comparable to a single wavelength unit cell, which we term “singular topological edge states” in this study.
The marked differences in wave localization between the BGs and the LRGs were experimentally verified using locally resonant granular-based metamaterials.
\section{Results and discussions}
\subsection{Extended SSH chain}
\begin{figure*}
    \includegraphics [width=0.98\textwidth]{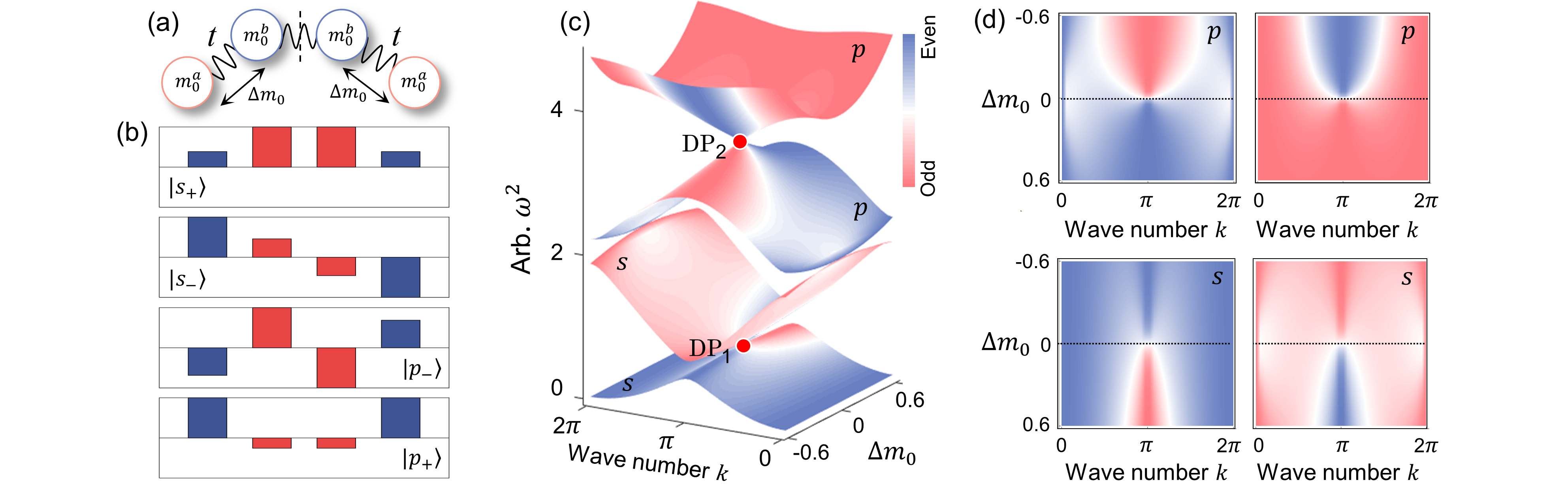}
    \caption{\textbf{Topological phase transition in the inversion-symmetric extended Su-Schrieffer-Hegger chain.} 
    (a) Representative mass-modulated ($\Delta{m}_0$) unit cell. (b) The mode profiles for different eigenmodes, characterized by two distinct symmetry operators. $s(p)$ represents the positive (negative) eigenvalues of the sublattice inversion symmetry operator $\mathcal{R}_s$ and $+(-)$ represents the positive (negative) eigenvalue of inversion symmetry $\mathcal{R}$. (c) Parametric band diagram as a function of mass difference ($\Delta{m}_0$=~$m_0^b-m_0^a$). 
    The color index indicates the parity of eigenmodes, which is obtained from the eigenvalues of the inversion operator. (d) The first two $s$ bands, comprise the lower Dirac point (DP1), and the first two $p$ bands, comprising the upper Dirac point (DP2). When $\Delta{m}_0$ becomes positive (negative), the parities between $p(s)$ bands start to mix, which indicates a topological phase transition.}
    \label{fig2}
\end{figure*}

In what follows, we introduce an extended SSH spring-mass chain without local resonance [see Fig.~\ref{fig2}(a)], which is an intermediate model toward our proposed model [Fig.~\ref{fig1}(c)].
The construction of an extended unit cell by connecting mirror-symmetric mass dimer pairs is related to the preservation of inversion symmetry.
The inversion symmetry plays a major role in achieving a nontrivial topology~\cite{Jiao_2021,Longhi_2018} although chiral symmetry is broken by mass dimerization.
In the extended SSH model, each sublattice mass ($m_0^{a(b)}$) is connected by constant stiffness $t$.
By invoking the Bloch-Floquet theory, the equations of motion for our unit cell yield an eigenvalue problem 
\begin{equation}
\left[\mathcal{H}_s(k)+\omega^2\mathcal{H}_m\right]\ket{\bm{\hat{u}}}=0.
\end{equation}
Here, $\mathcal{H}_m$ and $\mathcal{H}_s$ are Hermitian matrices composed of mass and spring constants, respectively (see Supplementary Note 1 for details); $\ket{\bm{\hat{u}}}=[u_1~u_2~u_3~u_4]^{\rm{T}}$ is the displacement vector of each site; $k$ is the wavenumber.
The bulk Hamiltonian $(\mathcal{H}$=~$\mathcal{H}_m^{-1}\mathcal{H}_s)$ conserves inversion ($\mathcal{R}$) symmetry, i.e., $\mathcal{R}\mathcal{H}(k)\mathcal{R}^{-1}$=~$\mathcal{H}(-k)$; where the inversion symmetry can be defined by the unitary operator $\mathcal{R}$=~$\sigma_x\otimes\sigma_x$.
Therefore, since this model exhibits inversion symmetry, we can define $\mathbb{Z}_2$ topological invariant $\nu^{n}$ as $(-1)^{\nu^{n}}$=~$\xi_0^n\xi_\pi^n$=~$\pm1$~\cite{Fu_2007}, where $n$ is the band index and $\xi_{k}$ is the eigenvalue of the inversion operator at the time-reversal symmetric point of the Brillouin zone.
\\
\indent
To characterize the shape of the eigenmodes in terms of symmetry (namely, orbital), we introduce a sublattice inversion operator $\mathcal{R}_s=\rm{I}\otimes\sigma_x$, switching positions between two sublattices $m^a_0$ and $m^b_0$.
Although the Hamiltonian does not commute with $\mathcal{R}_s$, this approach can be effective for defining eigenmodes.
In Fig.~\ref{fig2}(b), we shows the eigenmode profiles of $\ket{s_\pm}$ and $\ket{p_\pm}$ obtained from our systems. 
The $s(p)$ represents positive (negative) eigenvalues of $\mathcal{R}_s$ and the $+(-)$ index is the eigenvalue of inversion symmetry.
One can find that $s(p)$ orbital has in-phase (out-of-phase) displacement between the $a 
$ and $b$ sublattices.
\\
\indent
Next, we calculate a parametric band diagram as a function of the mass difference $\Delta{m}_0$=~$m_0^b-m_0^a$ [see Fig.~\ref{fig2}(c)].
In Fig.~\ref{fig2}(d), we present each energy band individually.
The lower two bands exhibit $s-$orbital characteristics, while the upper two bands exhibit $p-$orbital characteristics. 
The color index indicates the parity of the eigenmodes (blue for even, red for odd), determined by the eigenvalues of the inversion operator.
In the parametric band diagram, we observe well-defined Dirac points (DPs) at momentum $k=\pi$ when $\Delta{m}_0=0$.
These two DPs (for the lower and upper bands) arise from the degeneracy between each $s$ and $p$ band. 
Therefore, the topological phase transition, characterized by mixed parity, occurs in each of the $s$ and $p$ bands as 
$\Delta{m}_0$ deviates from zero.
Specifically, when $\Delta{m}_0>0$, the topological phase transition occurs in the lower $s$ bands, and when $\Delta{m}_0<0$, it occurs in the upper $p$ bands.
\\
\indent
To gain a deeper understanding of the underlying physics of our systems, we employ a perturbative approach to derive the effective Hamiltonian near the DPs.
Through a unitary transformation, we can map it to the well-known form of a massive Dirac Hamiltonian $\hat{\mathcal{H}}$~\cite{1928} (see Methods for the detailed derivation).
\begin{equation}
\begin{aligned}
    \hat{\mathcal{H}}(k)&=\frac{\sqrt{2}}{8{m_0}\omega_0}
    \begin{bmatrix}
    -\eta\Delta{{m_0}\omega_0^2} & i{t}\Delta{k} \\
    -it\Delta{k} & \eta\Delta{{m_0}\omega_0^2}
    \end{bmatrix}\\
    &=\frac{\sqrt{2}}{8{m_0}\omega_0}
    \left(t\Delta{k}\sigma_y\pm\Delta{{m_0}\omega_0^2}\sigma_z\right),
\end{aligned}
\end{equation}
where $\omega_0$ is the eigenfrequency at degeneracy, with $\eta=\mp{1}$ for $\text{DPs}_{1,2}$.
In this analogy, the diagonal element of the Hamiltonian―often called the Dirac mass―depends on the sign of $\Delta{m_0}$ and $\eta$.
For the topological phase, our attention is on the off-diagonal element―Dirac velocity―exhibiting odd behavior in momentum space.
This indicates coupling between different parities and is consistent with the results of the parametric band diagram.
Hence, despite the breaking of the well-known chiral symmetry, the inversion symmetry achieved through a modulated mass distribution in the unit cell enables the realization of a topologically nontrivial phase.
\subsection{Locally resonant system}
\begin{figure*}[htp]
    \includegraphics [width=0.98\textwidth]{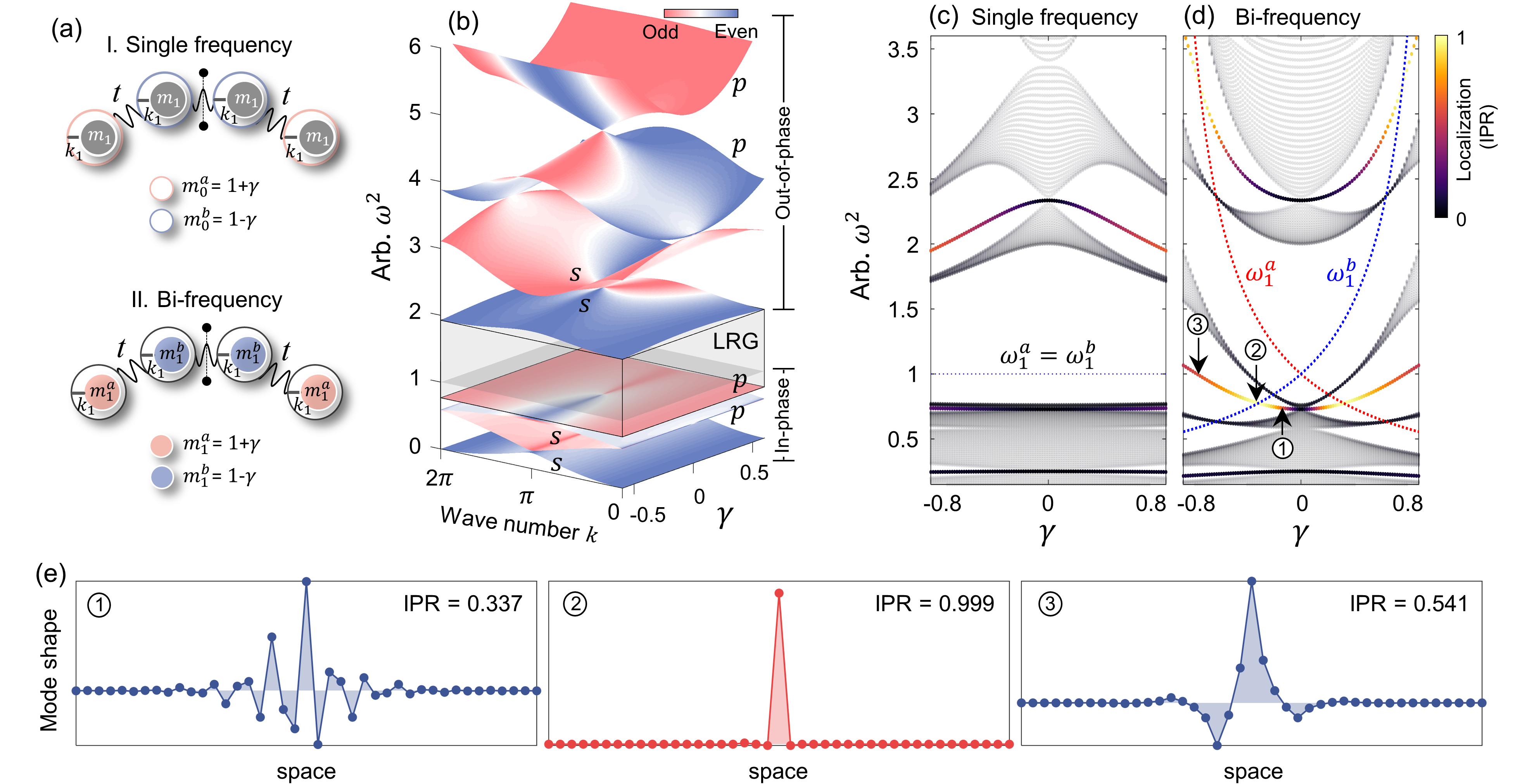}
    \caption{\textbf{Realization of singular topological edge state in locally resonant metamaterials.} (a) Schematic of the locally resonant unit cells for the single-frequency case (top) and the bi-frequency case (bottom), both preserving inversion symmetry. (b) Parametric band diagram in the single-frequency configuration. The four bands below the local resonance gap (LRG) exhibit in-phase motion between the host mass and local mass, while the four bands above the gap exhibit out-of-phase motion. (i.e., simple doubling of the bands). (c)-(d) Spectrum of finite local resonant systems, including bulk modes and topological edge states, as a function of local mass detuning ($\gamma$), where color denotes the localization index. The dotted lines represent the changes in local resonance frequency for each case. (c): In the single-frequency case, the topological edge states always reside within the BGs rather than the LRGs. The type of gap can be identified through the imaginary part of the band structure (not shown here). (d): In the bi-frequency case, the topologically nontrivial bandgaps transition from a BG to an LRG as $\gamma$ increases (third and fifth bandgap). Further increasing $\gamma$ leads to a reversion from an LRG back to a BG. (e) Evolution of the topological edge states at specific detuning values of local mass $\gamma$, as indicated by the black arrow in (d). When the topologically nontrivial bandgap becomes an LRG, the localization index (IPR) approaches nearly unity.}
    \label{fig3}
\end{figure*}

Next, we introduce an internal resonator while maintaining the inversion-symmetric configurations, as shown in Fig.~\ref{fig3}(a).
This mass-in-mass model effectively represents the subwavelength local resonance behavior of metamaterials~\cite{Ma_2016}. 
The internal resonators in the host mass have a coupling constant $k_1$ and a local mass $m_1$.
In this scheme, the effective mass of the unit cell, expressed as $m_\text{eff}(\omega)=m_0+m_1{\omega_1^2}/({\omega_1^2-\omega^2})$, where $\omega_1=\sqrt{k_1/m_1}$ (see Supplementary Note 2).
The effective mass exhibits frequency-dependent behavior, and as the frequency approaches the local resonant frequency, the effective mass diverges.
\\
\indent
Specifically, we investigate both \textit{single} and \textit{bi}-local resonance frequency scenarios, while keeping the hopping parameter $t$ and the local stiffness of the resonators $k_1$ constant.
In Fig.~\ref{fig3}(b), we show the parametric band diagram for a single frequency case related to $\gamma$, with the host mass $m_0$ denoted as $1$+$\gamma$ and $1$-$\gamma$. 
Introducing local resonators into the sublattice results in an avoided crossing and opens an LRG at $\omega_1$ (see the gray box), which expands the four bands to eight, i.e., doubling the bands.
The only difference between the upper four bands and the lower four bands, relative to the LRG, lies in the relative motion between the host mass and the local resonators. 
Specifically, the lower four bands exhibit in-phase motion between the host and local masses, while the upper four bands exhibit out-of-phase motion.
Thus, in the four bands of the subwavelength regimes ($\omega$$<$~$\omega_1$), we observe a similar mode parities pattern to that of the previous inversion-symmetric SSH chain [Fig.~\ref{fig2}(c)].
However, the LRG still remains topologically trivial regardless of the value $\gamma$ because there is no occurrence of a gap-closing and reopening process (i.e., a topological phase transition). 
This results in a conclusion similar to the subwavelength BGs achieved in many previous studies, while the topological states within LRGs remain unrealized.
In Fig.~\ref{fig3}(c), we present the spectrum for chains with different topologies when connected, highlighting the topological interface modes.
The color indicates the wave localization characteristics, which we quantify using the inverse participation ratio (IPR), defined as $({\sum_{n=1}^{N}u^4_n})/{(\sum_{n=1}^{N}u^2_n)^2}$. 
When all the energy is confined to a single particle, IPR=1. As the wave disperses, IPR approaches $1/N$, where $N$ is the total number of particles in the chain.
The dotted lines represent the local resonance frequencies.
We observe the emergence of topological edge modes within each classical BG, rather than in the LRG.
\begin{figure*}
    \includegraphics [width=0.98\textwidth]{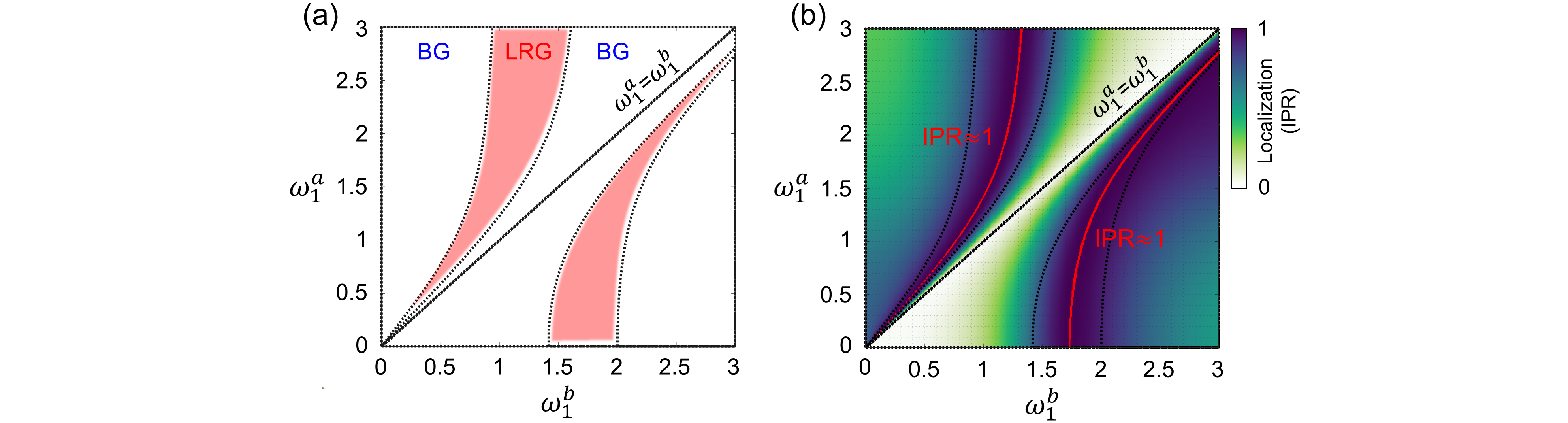}
    \caption{\textbf{Topological phase map and localization index map.} (a) Boundaries obtained by calculating phase maps in the parametric space of the bi-frequency resonance. This map describes the type of topologically nontrivial bandgap, indicating whether it is a BG or an LRG. The red areas highlight regions where topologically nontrivial bands undergo orbital transitions, i.e., from BG to LRG. (b) Localization index map in the same parametric space of the bi-frequency resonance. This map shows a significant increase in energy localization as the topologically nontrivial bandgap transitions to an LRG, with the red line indicating regions where the IPR equals 1.}
    \label{fig4}
\end{figure*}
\\
\indent
Moving on, in the bi-frequency case, the unit cell consists of a homogeneous host mass ($m_0$) with two different types of local masses ($m_1^{a,b}$).
This results in the formation of two distinct local resonance frequencies that vary with $\gamma$ (i.e., $\omega^a_1$=$1/\sqrt{1+\gamma}$ and $\omega^b_1$=$1/\sqrt{1-\gamma}$).
Fig.~\ref{fig3}(d) shows the calculated spectrum for the bi-frequency case.
With a gradual variation in $\gamma$, an intriguing observation unfolds: local resonance frequencies ($\omega_1^{a,b}$) begin to \textit{penetrate} into a topologically nontrivial bandgap, accompanied by a notable increase in the wave localization.
This indicates that the topological bandgap has transitioned from a classical BG to a strong LRG. 
Indeed, this bandgap transition is verified in the same manner by examining the imaginary part of the band structure derived from an infinite system (not shown here).
In addition, this gap transition is accompanied by an orbital transition, i.e., a transformation of the host mass's orbital from $p-$orbital to $s-$orbital, and vice versa.
This occurs because the relative motion of the host mass changes with respect to the local resonance gap.
\\
\indent
In Fig.~\ref{fig3}(e), we summarize the mode profiles of specific topological edge states indicated by the arrows in Fig.~\ref{fig3}(d).
The left and right panels depict the topological edge states within the BG, while the middle panel shows the edge states when the bandgap transitions to a LRG.
We observe that singular edge states emerge when the topological bandgap becomes an LRG. 
This is highlighted by an IPR of almost 1, indicating that most of the energy is confined within a subwavelength unit cell.
We attained this outcome by \textit{tuning} the local resonance frequency.
It should be noted that nontrivial topology in 1D systems is largely based on chiral symmetry originating from classical SSH chains. 
In this framework, tuning local resonance clearly breaks chiral symmetry, resulting in trivial topology.
This has, until now, hindered the realization of topological edge states in LRGs.
However, we have overcome this challenge by achieving a well-defined nontrivial topology through an inversion-symmetric configuration.
\\
\indent
The distinct difference in wave localization between the BG and LRG is explained by the differences in their bandgap mechanisms.
The gap properties of BG can be traced back to the work in Ref.~\cite{Allen_2000}. 
For an in-gap mode within the BG, the localization ($\mathcal{P}$) is determined by the ratio of two dimerization parameters, represented as $\mathcal{P}$~$\propto$~$\text{ln}(c_a/c_b)$.
In our system, the $c_{a,b}$ can be either mass or stiffness. 
According to this formula, achieving highly localized waves—where energy is confined to just a few unit cells—within this BG is impossible because the physical coefficient of $c_{a,b}$ must be either zero or infinite.
In contrast to the BG, highly localized waves can be successfully realized within the LRG.
According to the discussion in Ref.~\cite{Huang_2009}, the displacement of the particles within the local resonance gap takes the form $\rm{U}$~$\propto$~$e^{-\beta{x}}$, where $\beta$ is the imaginary part of the wavenumber and $x$ is space.
In this formula, if $\beta$ becomes infinite, most of the energy becomes highly localized to scales comparable to the unit cell.
At the local resonance frequency, the imaginary part of the wavenumber can indeed become infinite, and this condition can be achieved by tuning the local resonance frequency.
\begin{figure*}
    \includegraphics [width=0.98\textwidth]{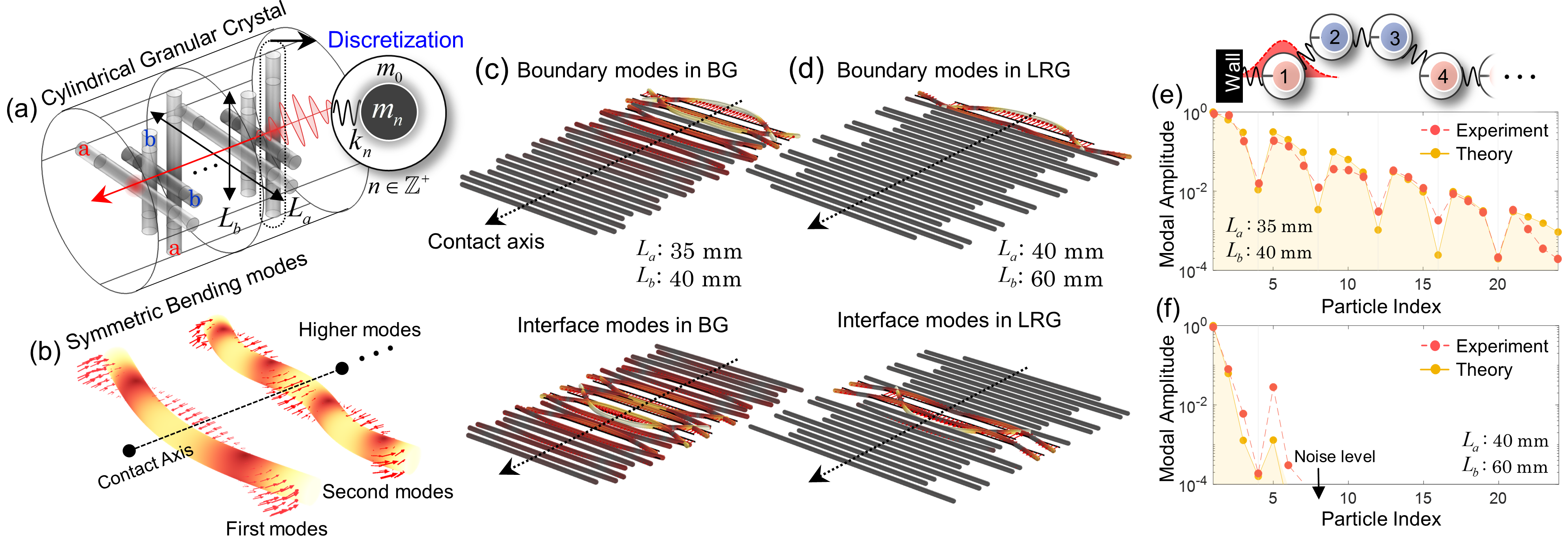}
    \caption{\textbf{Experimental observation of topologically protected singular edge states in a locally resonant granular-based metamaterial.} (a) Schematic of the experimental platform for a granule-based metamaterial, featuring two beams of different lengths ($L_a$ and $L_b$) that preserve an inversion-symmetric configuration. The continuum unit cell  (a single beam) is directly mapped to a spring-mass unit cell with multiple internal resonators using an analytical physic-informed discrete element modeling~\cite{Jang_2024}. 
    (b) Mode shape of bending vibrational modes of a cylindrical element, which exhibits strong coupling to external excitation (i.e., the local resonance effect). 
    (c)-(d) Numerical mode profiles of topological edge states: (c) within the BG and (d) within the LRG. To simplify the visualization of the mode profiles, the angular variation of adjacent cylinders is omitted; the dotted lines represent the actual contact axis. (e)-(f) Measured mode profiles of the topological boundary edge states within the BG and LRG (logarithmic scale of magnitude). The extracted experimental data (red markers) match the results of the eigenanalysis (yellow markers).}
    \label{fig5}
\end{figure*}
\subsubsection{Topological phase map}
We now examine a general framework for understanding how the topological bandgap transitions between a BG and an LRG depending on the local resonance conditions.
To characterize the topological phase together with the type of bandgap, we introduce a physical quantity $\Delta{\Phi}\equiv$~$\Phi_0$$-$$\Phi_\pi$, where $\Phi_{k}$ is $\mathrm{sign}(u_1/u_3)$.
This quantity allows us to consider not only the inversion eigenvalue, directly linked to the topological invariant but also the orbital information.
Thus, this quantity provides evidence for the bandgap transition.
Fig.~\ref{fig4}(a) shows the boundary line obtained by calculating $\Delta{\Phi}$ for the entire band structure in the parametric space of the bi-frequency resonance (see Supplementary Note 2.~B for details).
Within the boundary map obtained from the topological phase, the red areas represent regions where topologically nontrivial bands undergo orbital transitions.
\\
\indent
To validate this approach, we calculate the localization index (IPR) by focusing exclusively on topological gap states in finite chains, within the same parameter space illustrated in Fig.~\ref{fig4}(b).
In agreement with the phase map findings, we observe a significant increase in the localization index corresponding to the transition from the BG to the LRG. 
Specifically, the IPR approaches 1 near the centerline of the LRG regimes, indicating that energy is confined within a single unit cell, i.e., singular edge states.
As another approach, the spectral ordering, which can be intuitively examined, is investigated in Supplementary Note 2. C, demonstrating the identical bandgap transition condition.
\subsection{Experimental realization}
The remaining task is to experimentally demonstrate the transition of the topological bandgap from the classical BG to the LRG.
This can be achieved by comparing the differences in wave localization of the topological edge states between the BG and LRG.
To implement our theoretical framework, we need to map the structural elements of the locally resonant metamaterial onto the proposed discrete spring-mass model. Additionally, achieving tunable local resonance within each unit cell is essential.
Moreover, achieving tunable local resonance within a unit cell is crucial.
\\
\indent
To that end, we utilize locally resonant granular crystals known as woodpile metamaterials~\cite{Chong_2017,Kim_2015,Kim_2014}, composed of slender cylindrical beams [see Fig.~\ref{fig5}(a)].
In this testbed, the interaction potential between neighboring elements is governed by the Hertz contact law~\cite{Johnson1985contact}.
What distinguishes this system from other locally resonant metamaterials~\cite{Ma_2016,Liu_2000} is that the local resonance originates from the bending vibrational modes of the cylindrical elements themselves [see Fig.~\ref{fig5}(b)].
Thus, by adjusting the length of the cylinders, one can easily tune the local resonance frequency according to the relationship of $\omega\propto{1/L^2}$~\cite{inman1994}, where $L$ is the length of the beam.
As mentioned earlier, this continuous cylinder can be discretized using the recently developed physics-informed discrete element modeling~\cite{Jang_2024}.
This approach is rooted in analytic continuum beam theory and wave dynamics within periodic structures.
Therefore, with the geometry and material properties of the elastic cylindrical elements, the local resonance frequency can be determined and adjusted with analytic accuracy.
Detailed mathematical modeling is provided in Supplementary Note 3.
\\
\indent
Building on the discretized building blocks, we reapply and analyze the previously explored theoretical framework.
Through this approach, we chose the beam lengths $L_a$ and $L_b$ such that the topological edge states emerged in either the BG or LRG (i.e., the presence or absence of gap transition).
In addition, through finite element methods (see Method section), we further validate our experimental approach.
In Figs.~\ref{fig5}(c)-(d), we first show the numerical results of topological boundary modes and interface modes present in the BG and LRG, respectively.
To streamline the representation of the mode profile for topological edge states within the chain, the angular orientation of adjacent cylinders is omitted; the dotted lines indicate the actual contact axis.
We observe highly localized modes when edge states occur within the LRG, as opposed to the BG.
\\
\indent
In Figs.~\ref{fig5}(e)-(f), we show the experimentally extracted mode shapes (absolute values and logarithmic scale) for the topological boundary modes, alongside the theoretical results.
Details on the experimental setup and data acquisition are provided in the Methods section and Supplementary Fig.~S7.
The exponentially decaying profiles match closely between experiments and theory.
We demonstrate that energy localization of topological edge states is significantly greater in the LRG [Fig.~\ref{fig5}(f)] compared to the BG [Fig.~\ref{fig5}(e)], with a notable concentration in the initial part of the chain.
The modal amplitude of the remaining part of the chain shows noise from the experimental environment (see Supplementary Note 4).
More specifically, the localization index (IPR) is calculated as IPR = $0.427$ (theory) and $0.476 \pm 0.008$ (experiment) for the BG, and as IPR = $0.992$ (theory) and $0.975 \pm 0.006$ (experiment) for the LRG.
In this way, we demonstrate our topological singular edge states, which arise from the transition of the topological bandgap.
For the topological interface modes, we experimentally use input signals from the ends of a finite chain to couple with evanescent waves inside the band gap, which allows us to confirm their topological edge state indirectly. 
Related experimental validation is provided in Supplementary Note 5.
\section{Conclusion}
In this study, we demonstrate theoretically and experimentally herein the topological singular edge states in locally resonant metamaterials, leading to highly localized modes comparable to a subwavelength unit cell.
This singular state is achieved through a topological bandgap transition from a Bragg-type gap, induced by periodic modification, to a local resonance gap driven by strong coupling effects.
This bandgap transition can be achieved by tuning the local resonance frequency, which typically breaks chiral symmetry, but our study demonstrates a well-defined nontrivial topology through inversion symmetry.
Our findings open up possibilities for the extension of topological metamaterials by leveraging intrinsic bandgap properties, and we anticipate that this topological bandgap transition mechanism could provide insights for higher-dimensional topological systems e.g., highly localized surface states.
In addition, we believe our framework can be applied to various physical systems that support material or structural resonances, such as plasmonics, optics, and acoustics, among others.
\section{Methods}
{\itshape Effective Hamiltonian}.\xdash
We derive the effective Hamiltonian based on the perturbative approach. When the two types of host mass ($m_0^a, m_0^b$) become equal, the both lower and higher band gaps are closed and generate two lower and upper one-dimensional Dirac cones at the edge of the Brillouin zone. By the perturbation theory, the effective Hamiltonian of the perturbed system can be constructed by means of a set of eigenvectors that are obtained from the unperturbed Bloch Hamiltonian at the degeneracy point ($k=k_0$). 
At $k=k_0$, the generalized eigenvalue problem of an inversion symmetric spring-mass configuration [Fig.~2(a)] on the sublattice basis set $\begin{Bmatrix}
    \psi_1 \  \psi_2 \  \psi_3 \ \psi_4
\end{Bmatrix}$ is as follows:
\begin{equation}
\left[\mathcal{H}_s(k_0)+\omega_{0}^2\mathcal{H}_m\right]\ket{\bm{\hat{u}}_{k_0}^\pm}=0,
\end{equation}
where $\ket{\bm{\hat{u}}_{k_0}^\pm}$ is the degenerated eigenvector having the eigenvalue $\pm1$ for an inversion operator. Note that only $\mathcal{H}_s$ is dependent on Bloch momentum $k$, but not in the case of $\mathcal{H}_m$. By introducing the perturbation on $\mathcal{H}_m$ while conserving inversion symmetry, the above eigenvalue equation for the momentum near the degeneracy ($k=k_0+\Delta k$) is approximated as
\begin{equation}
\label{eq:eq}
    [\mathcal{H}_s(k_0+\Delta k)+\omega^2\mathcal{H}_m']\ket{\bm{\hat{u}}_k}=0
\end{equation}
Here,  $\mathcal{H}_s$ can be approximated as
\begin{equation}
    \mathcal{H}_s(k_0+\Delta k) \approx \mathcal{H}_k(k_0)+\mathcal{H}_k'(\Delta k),
\end{equation}
where the first-order approximation is applied as
\begin{equation}
\begin{split}
    \mathcal{H}_s'(\Delta k) & =\left(\frac{\partial{\mathcal{H}_k}}{\partial{k}}|_{k=k_0}\right)\Delta k \\ 
    & = i\Delta k\left(\ket{\psi_4}\bra{\psi_1}-\ket{\psi_1}\bra{\psi_4}\right).
\end{split}
\end{equation}And we write $\mathcal H_M'$ as:
\begin{equation}
\begin{split}
    \mathcal{H}_m' & =m_0^a(\ket{\psi_1}\bra{\psi_1}+\ket{\psi_4}\bra{\psi_4})+m_0^b(\ket{\psi_2}\bra{\psi_2}+\ket{\psi_3}\bra{\psi_3}) \\
    & =m\mathbb{I}+\frac{\Delta m_0}{2}\sigma_z\otimes\sigma_z,
\end{split}
\end{equation}
where $\Delta m_0\equiv m_0^a-m_0^b$ and $m_0\equiv({m_0^a+m_0^b})/{2}$.
In a perturbative limit, we can rewrite the $\ket{\bm{\hat{u}}_k}$ as the linear combination of degenerated eigenvectors as $\ket{\bm{\hat{u}}_{k}}=c_{+}\ket{\bm{\hat{u}}_{k_0}^{+}}+c_{-}\ket{\bm{\hat{u}}_{k_0}^{-}}$. Then, we can rewrite the generalized eigenvalue problem as
\begin{equation}
    \left[\mathcal{H}_k'(\Delta k)+\omega^2\mathcal{H}_m'-\omega_0^2\mathcal{H}_m\right]\ket{\bm{\hat{u}}_{k}}=0
\end{equation}
By approximating $\omega^2$ as $\omega_0^2+2\Delta{\omega}{\omega_0}$, above equation is represented as
\begin{equation}
\left[\mathcal{H}_k'(\Delta k)+\omega_0^2(\mathcal{H}_k'-\mathcal{H}_m)+2\omega_0\Delta{\omega}\mathcal{H}_m'\right]\ket{\bm{\hat{u}}_{k}}=0
\end{equation}
 Using the definition of $\mathcal{H}_m$ and $\mathcal{H}_k'$, equation becomes
 \begin{equation}
 -\frac{1}{2\omega_0m_0}[\mathcal{H}_k'(\Delta k)
 +\frac{\Delta m_0\omega_0^2}{2}(\sigma_z\otimes\sigma_z)] \ket{\bm{\hat{u}}_{k}}=\Delta{\omega}\ket{\bm{\hat{u}}_{k}}
\end{equation}
We can transform above equation with the transformation matrix $\mathcal{W} = \ket{\bm{\hat{u}}_{k_0}^{+}}\bra{\bm{\hat{u}}_{k_0}^{+}}+\ket{\bm{\hat{u}}_{k_0}^{-}}\bra{\bm{\hat{u}}_{k_0}^{-}}.$
Here, we newly define the two constituent matrices 
\begin{equation}
\begin{aligned}
\mathcal{H}_1 \equiv\mathcal{W}^\dag \mathcal{H}_k'(\Delta k)\mathcal{W}=\frac{\sqrt{2}}{4}t\Delta k
\sigma_y, \\
\mathcal{H}_2 \equiv\mathcal{W}^\dag \sigma_z\otimes\sigma_z\mathcal{W} =\eta\frac{\sqrt{2}}{2}\sigma_z,
\end{aligned}
\end{equation}
where $\eta$ is 1 for upper degeneracy and -1 for lower degeneracy; $t$ is spring constant and $\sigma_y$ is Pauli matrix for y component. The final form of the eigenvalue equation is expressed as follows:
\begin{equation}
    \mathcal{H}\psi=\Delta\omega\psi
\end{equation}
where $\psi=
\begin{bmatrix}
c_{+} \ c_{-}
\end{bmatrix}^\dag$. Here, the Bloch Hamiltonian becomes a massive Dirac Hamiltonian form represented by:
\begin{equation}
\begin{split}
    \mathcal{H}=-\frac{\sqrt{2}}{8m_0\omega_0}(t\Delta k \sigma_y \pm \Delta m_0 \omega_0^2 \sigma_z)\\
    =\frac{\sqrt{2}}{8m_0\omega_0}
    \begin{bmatrix}
    -\eta\Delta m_0\omega_0^2 & it\Delta k \\
    -it\Delta k & \eta\Delta m_0\omega_0^2
    \end{bmatrix}
\end{split}
\end{equation}
\vspace{2mm}
\\
{\itshape Numerical simulations}.\xdash 
Three-dimensional eigenfrequency simulations are conducted using the commercial finite element software, COMSOL Multiphysics. 
In the numerical simulations, 3D linear elasticity is applied to the cylindrical elements.
The interaction between neighboring cylinders is modeled as linear springs, with stiffness derived from linearized Hertzian contact (see Supplementary Note 3).
Following a methodology akin to the experiment, we utilized a total of 24 cylindrical beams of two different lengths ($L_a$ and $L_b$) in the chain configuration.
\vspace{2mm}
\\
{\itshape Experiments}.\xdash
Our elastic chain consists of 24 slender cylindrical elements made of fused quartz (Young’s modulus E=72~$\mathrm{GPa}$, Poisson’s ratio $\nu$=0.17, and density $\rho$=2187 $kg/m^3$) with low material damping. 
This allows us to observe the topological edge states in the cylinder chain with minimal wave dissipation.
To validate our predictions, we select two types of cylinder lengths ($L_a$ and $L_b$) using physics-informed discrete element modeling, ensuring the topological edge state emerges in either the BG or LRG.
This cylindrical element is orthogonally stacked in an inversion-symmetric configuration, with each cylinder having an identical diameter of 5 mm.
For a digital image of the experimental setup, see Supplementary Fig.~S7.
We then apply a static pre-compression force (18 $\mathrm{N}$) to the chain by placing a free weight, thus restricting the system dynamics to the linear regime under weak excitation amplitude.
At one end of the cylindrical chain, a piezo-actuator (PAHL 18$/$20) locally excites the first cylinder with a frequency sweep signal ranging from 1 to 50 kHz.
The particle velocity of each cylinder near the contact point is precisely measured using non-contact laser Doppler vibrometry (LDV), mounted on an automatic linear guide.
Reflection tape is affixed near the center of each cylinder to facilitate this measurement.
In this way, we scan the velocity profiles for all particles in the chain by moving the LDV.
We then obtain mode profiles experimentally by performing Fourier transformations on the temporal velocity profiles of all particles.
\bibliography{reference}
\section*{Acknowledgements}
This work was financially supported by the POSCO-POSTECH-RIST Convergence Research Center program funded by POSCO and the National Research Foundation (NRF) grant (RS-2024-00356928) funded by the Ministry of Science and ICT (MSIT) of the Korean government, and the grant (PES4400) from the endowment project of “Development of smart sensor technology for underwater environment monitoring” funded by Korea Research Institute of Ships Ocean engineering (KRISO). E.K. also acknowledges the support of the National Research Foundation grant (NRF-2020R1A2C2013414) and the Commercialization Promotion Agency for R\&D Outcomes(COMPA) grant (RS-2023-00304743) funded by the Korean Government (Ministry of Science and ICT, 2023).
\end{document}